\documentclass[aps,twocolumn,superscriptaddress,eqsecnum,longbibliography,floatfix]{revtex4-1}

\usepackage[T1]{fontenc}
\usepackage{newtxtext}
\usepackage{newtxmath}

\usepackage{amsmath}
\usepackage{mathtools}
\usepackage{amsfonts}
\usepackage{bm}

\usepackage{makecell}

\usepackage{color}
\usepackage{xcolor}
\usepackage[colorlinks,
            citecolor=blue,
            linkcolor=red,
            urlcolor=blue]{hyperref}

\usepackage{graphicx}
\usepackage[all]{hypcap} 

\begin{document}

\title{From trivial to topological paramagnets: The case of $\mathbb{Z}_2$ and $\mathbb{Z}_2^3$ symmetries in two dimensions}

\author{Maxime Dupont}
	\affiliation{Department of Physics, University of California, Berkeley, California 94720, USA}
	\affiliation{Materials Sciences Division, Lawrence Berkeley National Laboratory, Berkeley, California 94720, USA}
\author{Snir Gazit}
	\affiliation{Racah Institute of Physics and the Fritz Haber Center for Molecular Dynamics, The Hebrew University, Jerusalem 91904, Israel}
\author{Thomas Scaffidi}
	\affiliation{Department of Physics, University of Toronto, Toronto, Ontario, M5S 1A7, Canada}

\begin{abstract}
	Using quantum Monte Carlo simulations, we map out the phase diagram of Hamiltonians interpolating between trivial and nontrivial bosonic symmetry-protected topological phases, protected by $\mathbb{Z}_2$ and $\mathbb{Z}_2^3$ symmetries, in two dimensions. In all cases, we find that the trivial and the topological phases are separated by an intermediate phase in which the protecting symmetry is spontaneously broken. Depending on the model, we identify a variety of magnetic orders on the triangular lattice, including ferromagnetism, $\sqrt{3}\times\sqrt{3}$ order, and stripe orders (both commensurate and incommensurate). Critical properties are determined through a finite-size scaling analysis. Possible scenarios regarding the nature of the phase transitions are discussed.
\end{abstract}

\maketitle

\section{Introduction}

The Landau-Ginzburg-Wilson (LGW) paradigm for classifying phases of matter, based on their symmetry and its subsequent breaking~\cite{landau1937,sachdev2001}, has been challenged by the discovery of symmetry-protected topological (SPT) phases~\cite{Chen1604,Chen2011,Chen2011b,YuanMing2012,PhysRevB.91.134404,pollmann2010,pollmann2012}. Of particular interest are SPT phases that arise due to strong correlations, generalizing the free fermion band-structure classification of topological insulators to generic interacting systems. SPT phases sharing the~\textit{same} protecting symmetry display identical physical properties in the bulk and are indistinguishable by symmetry-based probes such as local order parameters. Instead, the distinction between trivial and nontrivial SPT phases is more subtle and manifests itself after gauging the protecting symmetry, or through properties like string order parameters, edge states, entanglement spectrum, strange correlators, etc.~\cite{Chen1604,Chen2011,Chen2011b,YuanMing2012,PhysRevB.91.134404,pollmann2010,pollmann2012,Cenke2014,ringel2015,PhysRevB.93.115105,PhysRevB.91.195134,PhysRevB.89.195122,PhysRevLett.114.031601}. Well known examples include the Haldane phase of odd-integer Heisenberg chains~\cite{haldane1983spin,haldane1983b,affleck1987,affleck1987aklt,affleck1988,affleck1989,PhysRevB.80.155131,pollmann2010,pollmann2012} and the bosonic integer quantum Hall phases in two dimensions~\cite{senthil2013,furukawa2013,wu2013b,regnault2013,grover2013,geraedts2013,he2015,geraedts2017}, among others.

While the topological classification of SPT phases is by now fairly well established~\cite{Chen1604,Chen2011,Chen2011b,YuanMing2012,PhysRevB.91.134404,pollmann2010,pollmann2012}, our understanding of quantum phase transitions involving them is still lacking. Such transitions are expected to give rise to novel quantum critical behavior, going beyond the LGW predictions. In that respect, phase transitions between SPT phases and the more familiar symmetry-broken, ordered states~\cite{PhysRevB.83.174409,grover2012quantum,PhysRevB.91.235309,parker2017,parker2018,parker2019,verresen2017,verresen2018,2019arXiv190506969V,PhysRevB.103.L100207} as well as transitions between trivial and nontrivial SPT phases~\cite{grover2013,lu2014,morampudi2014,tsui2015,tsui2015b,you2016,he2016,you2018,tsui2017,geraedts2017,bi2019,bultinck2019,gozel2019,zeng2020} have both attracted tremendous attention. Previous works have mostly considered transitions between SPT phases that are protected by continuous symmetries and uncovered remarkable relations with deconfined quantum criticality~\cite{Senthil1490,PhysRevX.7.031051,PhysRevX.7.031052,you2016,he2016,you2018,geraedts2017,bi2019,zeng2020}. However, the study of microscopic models with discrete symmetry groups has remained relatively scarce~\cite{morampudi2014,huang2016,tsui2017,verresen2017,PhysRevB.97.195124,xu2018}.
\begin{figure}[t]
	\center
	\includegraphics[width=1.0\columnwidth,clip]{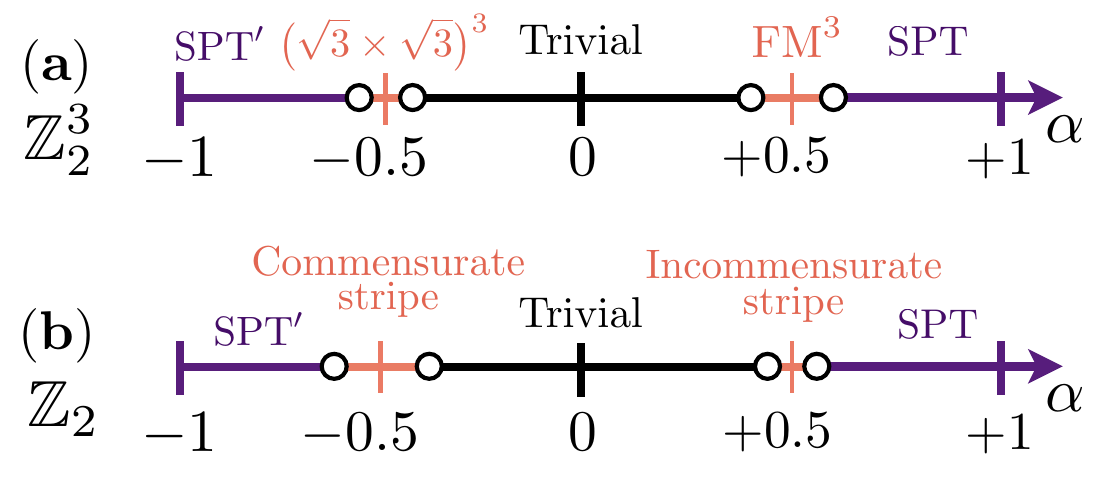}
	\caption{Quantum phase diagram for the Hamiltonians~\eqref{eq:hamiltonian}. For each symmetry ($\mathbb{Z}^3_2$ and $\mathbb{Z}_2$), the Hamiltonian interpolates between the parent Hamiltonian of a paramagnet at $\alpha=0$, the Hamiltonian of a nontrivial SPT at $\alpha=1$, and the Hamiltonian of another nontrivial SPT (called SPT') for $\alpha=-1$. The only difference between SPT and SPT' is a minus sign in front of the parent Hamiltonian, which confers a nontrivial weak $0$D index to SPT'. In all cases, the transition from the trivial to the SPT phases happens via an intermediate symmetry-breaking phase, where the protecting symmetry is spontaneously broken. The $\mathbb{Z}^3_2$ model at positive and negative $\alpha$ as well as the $\mathbb{Z}_2$ model for $\alpha<0$ are studied in this paper. The other case of the $\mathbb{Z}_2$ model with $\alpha>0$ was thoroughly studied in Ref.~\onlinecite{dupont2020}. See Table~\ref{tab:transitions} for a more quantitative summary of our findings.}
	\label{fig:phase_diagram}
\end{figure}

In the absence of an overarching theoretical framework of SPT criticality, one must resort to exact numerical methods to determine their properties in specific cases. In that regard, in dimensions $D>1$, exact diagonalization and density matrix renormalization group (DMRG) techniques~\cite{white1992,white1993} are restricted to small system sizes, which typically do not allow studies of long-wavelength universal properties. Also, in many cases, quantum Monte Carlo techniques are plagued by the numerical ``sign problem'' rendering the statistical errors uncontrolled~\cite{troyer2005}. Therefore, identifying concrete lattice models that exhibit transitions involving SPTs and that are amenable to an unbiased numerical solution is desirable.

In this paper, we numerically investigate the phase diagram of two different models that interpolate between trivial and topological paramagnets, protected by $\mathbb{Z}_2$ and $\mathbb{Z}_2^3$ symmetries, see also Ref.~\onlinecite{dupont2020}. Crucially, the interpolation does not explicitly break the protecting symmetry, and since it connects two distinct phases of matter, we expect a quantum phase transition to occur along the way. This can happen either through a single transition point separating the two phases or a two-step transition via an intermediate phase, which spontaneously breaks the protecting symmetry. In all cases studied in this work, we find that the latter scenario is realized, giving rise to an intermediate magnetically ordered phase, see Fig.~\ref{fig:phase_diagram}. Interestingly, in certain cases, magnetic order is accompanied by additional broken symmetries, such as lattice translations or point group symmetries.

The rest of the paper is organized as follows: In Sec.~\ref{sec:models}, building on several exactly solvable models of SPT phases \cite{levin2012,YOSHIDA2017387}, we construct a family of one-parameter Hamiltonians, which interpolate between trivial and topological SPT phases. In Sec. ~\ref{sec:method} we present a sign-problem free quantum Monte Carlo method used to numerically study these models and discuss the physical observables used to probe the various emergent phases and phase transitions. We present our numerical results and discuss their physical interpretation in Sec.~\ref{sec:results}. Lastly, we summarize our findings and highlight future research directions in Sec.~\ref{sec:conclusion}.

\section{Models }
\label{sec:models}

\begin{figure}[!t]
	\center
	\includegraphics[width=1.0\columnwidth,clip]{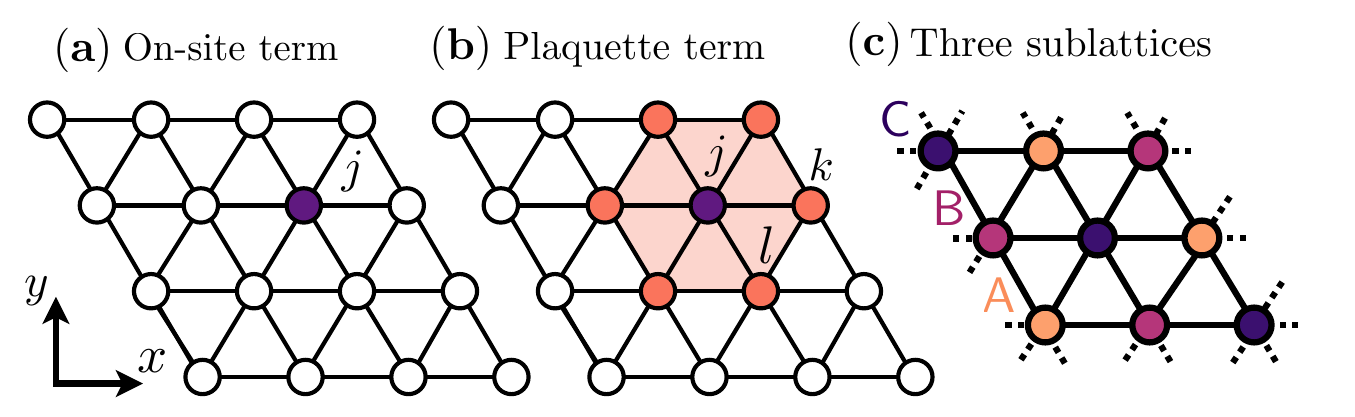}
	\caption{The Hamiltonian~\eqref{eq:hamiltonian} defined on the triangular lattice is made of two distinct parts. (\textbf{a}) The first one~\eqref{eq:ham_tri} describes a trivial gapped Ising paramagnet with onsite terms. (\textbf{b}) The second part~\eqref{eq:ham_spt} describes a topological gapped Ising paramagnet protected by the $\mathbb{Z}_2$ Ising spin-flip symmetry, with plaquette terms involving the six nearest neighbors of a given lattice site $j$. See Eqs.~\eqref{eq:plaquette_operator_Z23} and~\eqref{eq:plaquette_operator_Z2} for the microscopic definitions of the plaquette operator considered in this work. (\textbf{c}) Different sublattices for the $\mathbb{Z}_2^3$ model on the triangular lattice geometry considered in this work.}
	\label{fig:model}
\end{figure}

In this section we present two single-parameter Hamiltonians, admitting $\mathbb{Z}_2$ and $\mathbb{Z}_2^3$ symmetries, that interpolate between trivial and symmetry-protected topological phases. The degrees of freedom of our models are Ising spins residing on the sites of a two-dimensional triangular lattice, see Fig.~\ref{fig:model}. A trivial Ising paramagnet is simply defined by the Hamiltonian,
\begin{equation}
    \mathcal{H}_\mathrm{tri}=-\sum\nolimits_{j}\sigma^x_j,
    \label{eq:ham_tri}
\end{equation}
where $\sigma_j^{x,y,z}$ are the standard Pauli matrices ($\sigma_j^z=\pm 1$) defined on site $j$, see Fig.~\ref{fig:model}\,(a). The above Hamiltonian has a unique gapped ground state, given by the product state $|\Psi_{\mathrm{gs}}\rangle=\prod_j|\sigma^x_j=+1\rangle$ in the $\sigma^x$ basis.

On the same lattice, we can also define nontrivial SPTs with parent Hamiltonians taking the form~\cite{Chen1604,Chen2011,Chen2011b,levin2012,Yoshida,YOSHIDA2017387}:,
\begin{equation}
    \mathcal{H}_\mathrm{topo}=-\sum\nolimits_{j}\sigma^x_j\vartheta_j.
    \label{eq:ham_spt}
\end{equation}
Here, $\vartheta_j$ is a plaquette operator centered around site $j$ and involving all its neighbors, as sketched in Fig.~\ref{fig:model}\,(b). It is diagonal in the computational basis $|\sigma_1^z,\sigma_2^z,\ldots\sigma_N^z\rangle$ with eigenvalues $\pm 1$ . In a geometry with periodic boundary conditions and with appropriate choices of $\vartheta_j$, as we detail in the following subsections, the above Hamiltonian has a unique gapped ground state realizing a symmetry protected topological phase.

\footnotetext[2]{A system in $D$ spatial dimensions can have nontrivial ``weak'' $D'$-dimensional SPT order, for $D'<D$~\cite{Chen2011}. The protection of this weak SPT order requires translation symmetry to be respected, on top of the protecting on-site symmetry. For $D'=0$ and a $\mathbb{Z}_2$ symmetry, the classification is $\mathbb{Z}_2$. For a symmetry generated by $\sigma^x$, parent Hamiltonians of the two zero-dimensional SPT classes are given by $-\sigma^x$ and $\sigma^x$, and the twisting unitary which relates the two is given by $\sigma^z$.}

Since the ground states of~\eqref{eq:ham_tri} and~\eqref{eq:ham_spt} describe a trivial and nontrivial SPT phases, respectively, then a quantum phase transition is expected to occur along a symmetry-preserving path in parameter space that interpolates between the two. Note that the ground states of $\mathcal{H}_\mathrm{topo}$ and $-\mathcal{H}_\mathrm{topo}$ have the same strong SPT index, but a different weak zero-dimensional SPT index~\cite{Note2}.

This enables us to study two distinct transitions for each symmetry group, with the following Hamiltonian,
\begin{equation}
    \mathcal{H}(\alpha)=\Bigl(1-|\alpha|\Bigr)\mathcal{H}_\mathrm{tri}+\alpha\mathcal{H}_\mathrm{topo},
    \label{eq:hamiltonian}
\end{equation}
with $\alpha\in[-1,+1]$. The transition for $\alpha>0$ involves changing only the strong SPT index. The transition for $\alpha<0$  involves an additional change in the weak zero-dimensional SPT index. In the following we give explicit expressions for the plaquette operators $\vartheta_j$ of Eq.~\eqref{eq:ham_spt}.

\subsection{ $\mathbb{Z}_2^3$ model}

We first consider an SPT phase protected by a $\mathbb{Z}_2^3$ symmetry corresponding to flipping all the spins belonging to each one of the three sublattices of the triangular lattice, see Fig.~\ref{fig:model}\,(c). The associated generators of this symmetry are $G_{\mathsf{A}/\mathsf{B}/\mathsf{C}}=\prod_{i \in \mathsf{A}/\mathsf{B}/\mathsf{C}} \sigma^x_i$.

The plaquette operator is defined as~\cite{Chen2011,Yoshida,YOSHIDA2017387},
\begin{equation}
    \vartheta_j^\mathrm{(\mathbb{Z}_2^3)}=\prod\nolimits_{\bigtriangleup_{jkl}} (-1)^{\frac{1}{4}(1-\sigma^z_k)(1-\sigma^z_l)}.
    \label{eq:plaquette_operator_Z23}
\end{equation}
To evaluate the above product, one counts the number of nearest-neighbor spin pairs belonging to the plaquette surrounding $j$ and both taking the value $-1$. If the number of such pairs is odd the product equals $-1$, and otherwise it equals $+1$. With the above definition for $\vartheta_j^\mathrm{(\mathbb{Z}_2^3)}$, we can relate $\mathcal{H}_\mathrm{topo}^{(\mathbb{Z}_2^3)}$ to the trivial paramagnet, $\mathcal{H}_\text{tri}$, by the following unitary transformation,
\begin{equation}
    \mathcal{H}_\mathrm{topo}^{(\mathbb{Z}_2^3)} = \left(\mathcal{U}^{(\mathbb{Z}_2^3)}\right)^\dagger\;\mathcal{H}_\mathrm{tri}\;\mathcal{U}^{(\mathbb{Z}_2^3)},
    \label{eq:uni_trans_z23}.
\end{equation}
Here, $\mathcal{U}^{(\mathbb{Z}_2^3)}=(-1)^{N_\mathrm{\triangle_{---}}}$ is a diagonal (in the computational basis) unitary operator, where $N_\mathrm{\triangle_{---}}$ counts the number of triangles with three $-1$ spins in a given basis configuration. Importantly, $\mathcal{H}_\mathrm{topo}^{(\mathbb{Z}_2^3)}$ commutes with the  $\mathbb{Z}_2^3$ symmetry. The resulting SPT phase corresponds to a type-III cocycle and therefore couples nontrivially to the three different sublattice Ising symmetries. Gauging the $\mathbb{Z}_2^3$  symmetry gives rise to a non-Abelian $D_4$ quantum double phase~\cite{deWild,Yoshida,YOSHIDA2017387,PhysRevB.93.115136,PhysRevB.95.035131,PUTROV2017254}.

Since $\mathcal{U}^{(\mathbb{Z}_2^3)} =(\mathcal{U}^{(\mathbb{Z}_2^3)})^\dagger$, we find that for $\alpha\geq 0$, $\mathcal{H}(\alpha)$ is related by a unitary transformation to $\mathcal{H}(1-\alpha)$. Furthermore, since $-\mathcal{H}_\mathrm{topo}$ is related to $\mathcal{H}_\mathrm{tri}$ by a unitary transformation with unitary operator $\mathcal{U}^{(\mathbb{Z}_2^3)} \prod_{i}\sigma^z_i$, we also have a duality relating $\mathcal{H}(\alpha)$ to $\mathcal{H}(-1-\alpha)$ for $\alpha\leq 0$. In other words, the phase diagram is symmetric about $\alpha=0.5$ for $\alpha>0$ and around $\alpha=-0.5$ for $\alpha<0$.

The above relations are a key property of our model since, as we show below, in the computational $\sigma^z$ basis, the Hamiltonian is sign-problem free only within the parameter range $\alpha\in [-0.5,0.5]$. We can, therefore, determine all physical properties within this range by means of quantum Monte Carlo simulations and treat the rest of the phase diagram using the above duality relations.

\subsection{ $\mathbb{Z}_2$ model}

The second model we consider is an SPT phase protected by a $\mathbb{Z}_2$ symmetry corresponding to a global Ising spin flip, $\sigma^z_i\to-\sigma^z_i~\forall i$. In this case, the plaquette operator takes the form~\cite{Chen1604,Chen2011,Chen2011b,levin2012},
\begin{equation}
    \vartheta_j^{(\mathbb{Z}_2)}=\prod\nolimits_{\bigtriangleup_{jkl}} i^{\frac{1}{2}(-1-\sigma^z_k\sigma^z_l)}.
    \label{eq:plaquette_operator_Z2}
\end{equation}
The above product gives a minus sign if the number of nearest neighbor spin pairs belonging to the plaquette surrounding $j$ and pointing in the same direction equals to $2$ or $6$, and gives a plus sign otherwise. Such pairs can only come in even numbers ($0,2,4,6$), such that $\vartheta_j^{(\mathbb{Z}_2)}$ is a Hermitian operator despite the presence of the imaginary number $i$ in its definition. $\mathcal{H}_\mathrm{topo}^{({\mathbb{Z}_2})}$ is related to the trivial paramagnet $\mathcal{H}_\mathrm{tri}$  by the following unitary transformation,
\begin{equation}
    \mathcal{H}_\mathrm{topo}^{({\mathbb{Z}_2})} = \left(\mathcal{U}^{({\mathbb{Z}_2})}\right)^\dagger\;\mathcal{H}_\mathrm{tri}\;\mathcal{U}^{({\mathbb{Z}_2})}.
    \label{eq:uni_trans_Z2}
\end{equation}
Here, $\mathcal{U}^{({\mathbb{Z}_2})}=(-1)^{N_\mathrm{DW}}$, where $N_\mathrm{DW}$ counts the number of domain walls in a given spin configuration. The number of domain walls is well-defined on the triangular lattice since each spin configuration defines a unique configuration of closed and nonintersecting domain walls on the dual honeycomb lattice. We note that $\mathcal{U}^{({\mathbb{Z}_2})}$ is diagonal in the computational basis. Gauging the Ising symmetry in this model realizes the double semion phase~\cite{levin2012}.

By the same argument as in the $\mathbb{Z}_2^3$ case, we find a duality relating $\alpha$ and $(1-\alpha)$ for $\alpha\geq 0$, and relating $\alpha$ and $(-1-\alpha)$ for $\alpha\leq 0$. As before, owing to this symmetry, and we can limit ourselves to calculations in the range $\alpha\in [-0.5,0.5]$ using a sign problem free quantum Monte Carlo simulations and obtain results for the rest of the phase diagram using the above duality relations.

\section{Numerical Methods: Quantum Monte Carlo and observables}
\label{sec:method}

\subsection{Stochastic Series expansion}

For convenience, and without loss of generality, we first rewrite the Hamiltonian~\eqref{eq:hamiltonian} as
\begin{equation}
    \mathcal{H} = \sum\nolimits_{j=1}^N X_j\quad\mathrm{with}\quad X_j=-\sigma^x_j\Bigl(1-|\alpha|+\alpha \vartheta_j\Bigr),
    \label{eq:hamiltonian_qmc}
\end{equation}
with $N$ being the total number of lattice sites.  In what follows, we always impose periodic boundary conditions.
\begin{figure}[!t]
	\center
	\includegraphics[width=0.8\columnwidth,clip]{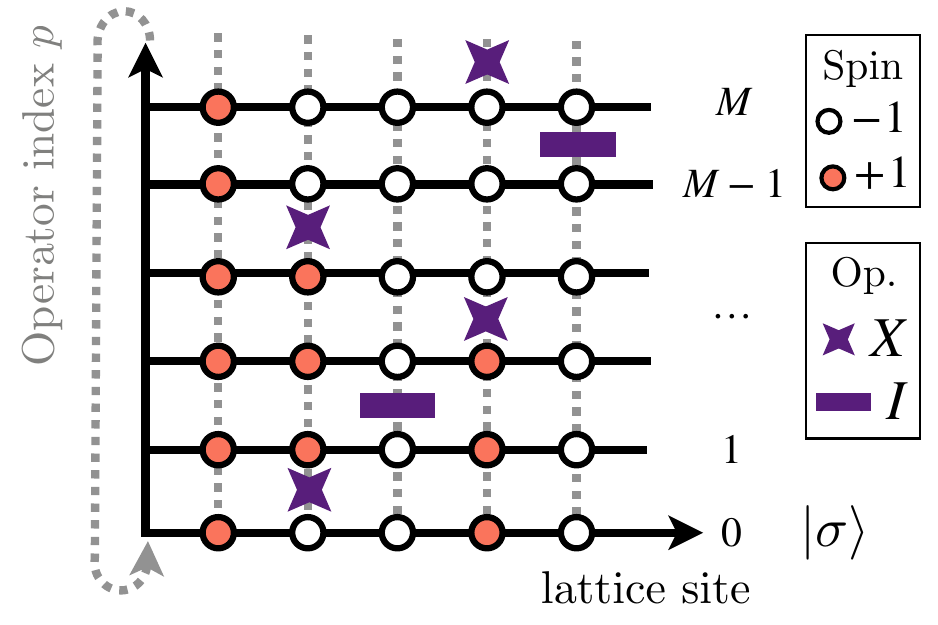}
	\caption{Example of a quantum Monte Carlo configuration for the model~\eqref{eq:hamiltonian_qmc} in the SSE formulation. It consists of a basis state $|\sigma\rangle\equiv|\sigma_1\sigma_2\ldots\sigma_N\rangle$ (at index $p=0$) and a sequence $S_M$ of operators of fixed length $M$. There are only two types of operators in our case: Identity $I$ and spin-flip $X$. Here, the identity operators have been assigned a lattice site. The algorithm aims at sampling the configuration space $\{|\sigma\rangle, S_M\}$.}
	\label{fig:qmc_conf}
\end{figure}

Within the stochastic series expansion (SSE) formulation of quantum Monte Carlo, the partition function of the system at inverse temperature $\beta=1/T$ reads~\cite{sandvik2010,sandvik2019},
\begin{equation}
    \mathcal{Z}=\sum_{\{|\sigma\rangle\}}\sum_{\{S_M\}}\frac{(-\beta)^n(M-n)!}{M!}\Bigl\langle\sigma\Bigl| \prod_{p=0}^M O(p) \Bigr|\sigma\Bigr\rangle,
    \label{eq:partition_function}
\end{equation}
where the configuration space is defined by all possible combinations of basis states $\{|\sigma\rangle\equiv|\sigma_1,\sigma_2,\ldots\sigma_N\rangle\}$ and sequence of operators $\{S_M\}$ of fixed length $M$. For a given sequence of operators $S_M$, the operators are denoted by $O(p)$ with position index $p$. They can either be the identity $I(p)$ or $X(p)$, as defined in Eq.~\eqref{eq:hamiltonian_qmc}, see Fig.~\ref{fig:qmc_conf}. Spin-flip operators $X(p)$ act on lattice sites $j\in[1, N]$ but its index can be omitted because it is actually redundant with the sequence $S_M$ considered, unless specified otherwise. The integer $n$ is the number of nonidentity operators in the sequence $S_M$. Note that $M$ should be taken large enough such that $n<M$ is ensured in practice. One can rewrite the partition function~\eqref{eq:partition_function} as
\begin{equation}
    \mathcal{Z}=\sum_{\{|\sigma\rangle\}}\sum_{\{S_M\}}W\Bigl(|\sigma\rangle, S_M\Bigr),
\end{equation}
where $W(|\sigma\rangle, S_M)$ is the weight of a configuration with a probability $P(|\sigma\rangle,S_M)=W(|\sigma\rangle, S_M)/\mathcal{Z}$. We want to sample these configurations in a Monte Carlo fashion, which supposes that $W(|\sigma\rangle, S_M)\geq 0$ for all configurations, otherwise we end up with the infamous sign problem~\cite{troyer2005}. Because the number $n$ of spin-flip operators~\eqref{eq:hamiltonian_qmc} is even to respect the periodicity along the ``operator index axis'' (see Fig.~\ref{fig:qmc_conf}), this condition is fulfilled for all the models considered in the range $\alpha\in[-0.5, +0.5]$.

There are no known efficient loop or cluster-type updates~\cite{sandvik1999,syljuasen2002,sandvik2003,evertz2003} for the models~\eqref{eq:hamiltonian_qmc}, and we can only rely on local moves in the configuration space~\cite{sandvik1992}. This limits the system sizes one can access in practice to a few hundred lattice sites. Assuming some valid configuration is defined by $\{|\sigma\rangle, S_M\}$, the updates that we propose involve changes in the sequence of operators $S_M$ that will indirectly involve changes in the basis state $|\sigma\rangle$. There cannot be two operators with the same index $p$, and there can only be an even number of $X$ operators at a given lattice site $j$; otherwise, the two states $|\sigma\rangle$ sandwiching the product of operators in~\eqref{eq:partition_function} would be different. Full details on the algorithm implementation are discussed in Appendix~\ref{app:qmc_sse}.

Because of the constraints in the different models as $\alpha\to\pm 1/2$ (some spin flips are strictly prohibited at $\alpha=\pm1/2$, as explained below), we have found that the SSE algorithm with local updates gives incorrect results when one gets very close to $\alpha=\pm 1/2$ (by comparing to exact diagonalization on small system sizes). We believe this is an ergodicity issue in the SSE configuration space due to the nature of our updates. Note that this problem does not concern the data shown in this work since they are relatively far away from $\alpha=1/2$. However, a study of the $\mathbb{Z}_2$ model at $\alpha=1/2$ was necessary in Ref.~\onlinecite{dupont2020}. We have, therefore, developed a complementary algorithm specifically to study that case. This algorithm is based on projective quantum Monte Carlo~\cite{becca2017} and does not suffer from ergodicity issues, see Appendix~\ref{app:qmc_proj}.

\subsection{Physical observables}

To determine the different quantum phases and phase transition in each model, we focus primarily on the spin structure factor associated with the imaginary-time two-point correlation function,
\begin{equation}
    C\bigl(\boldsymbol{q},\tau\bigr)=\frac{1}{N}\sum\nolimits_{\boldsymbol{r}}\mathrm{e}^{-i\boldsymbol{q}\cdot\boldsymbol{r}}\Bigl\langle\sigma^z_{\boldsymbol{r}}(\tau)\sigma^z_{\boldsymbol{0}}(0)\Bigr\rangle.
    \label{eq:zz_corr}
\end{equation}
This quantity is readily computed in SSE simulation, since it is diagonal in the computational basis~\cite{sandvik1991,sandvik1997,sandvik2010}. The equal time correlation function ($\tau=0$) probes potential spontaneous magnetic order marked by a peak at an ordering wave vector $\boldsymbol{q}\equiv\boldsymbol{q}_\mathbf{order}$. The gap $\Delta$ between the two $\mathbb{Z}_2$ Ising symmetric sectors is also accessible by examining the long imaginary time asymptotic decay \begin{equation}
    C(\boldsymbol{q}_\mathbf{order},\tau)\propto\exp(-\Delta\tau)
    \label{eq:zz_corr_tau}
\end{equation}
with $\tau\in[0,\beta/2]$. Periodic systems of finite size $N=L\times L$ are considered, with the lattice geometry of Fig.~\ref{fig:model}. We set the inverse temperature of the SSE algorithm at $\beta=4L$, which we found to be sufficiently low to probe the ground state of the models studied.

\section{Numerical results}
\label{sec:results}

\begin{figure*}[!t]
	\center
	\includegraphics[width=2.0\columnwidth,clip]{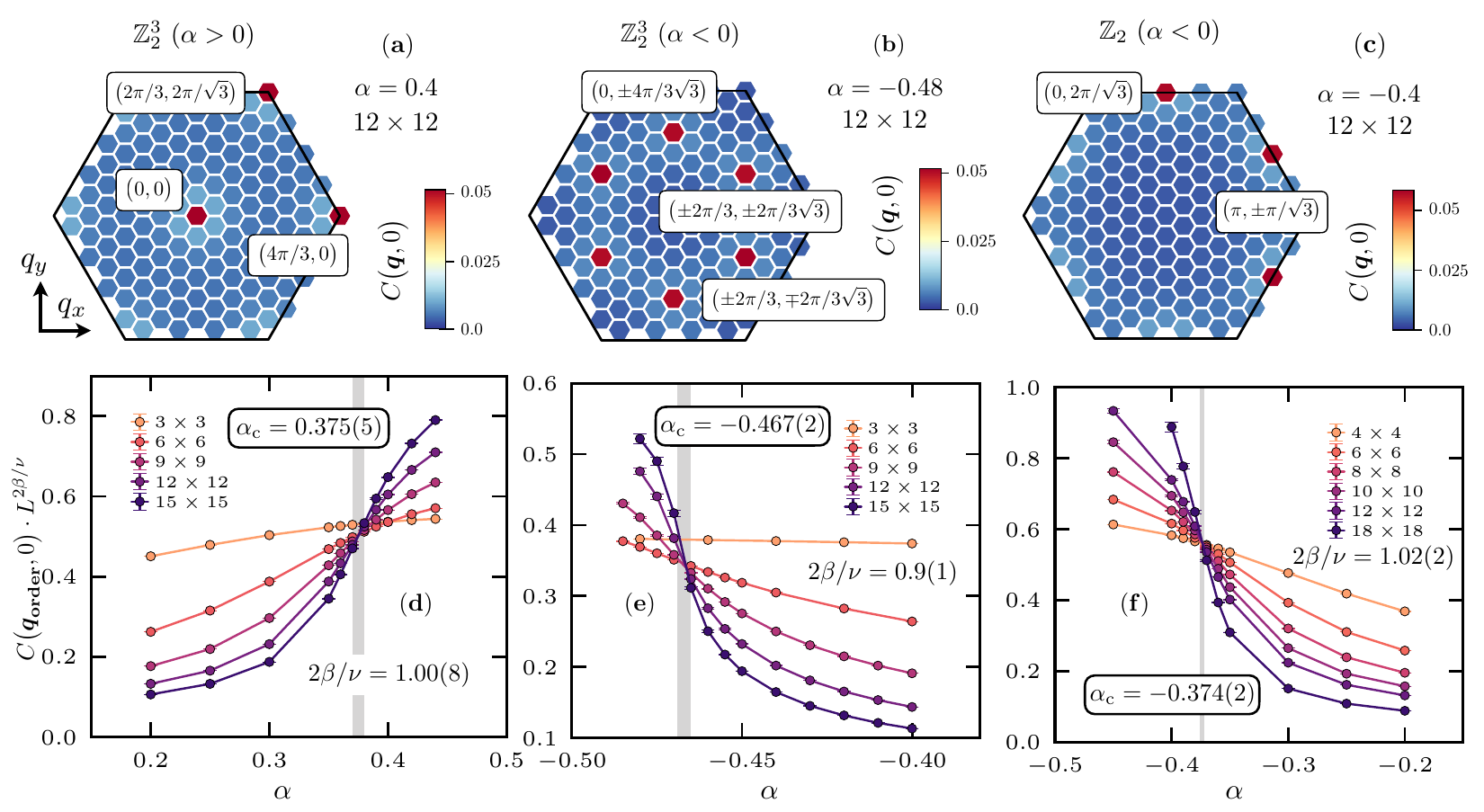}
	\caption{(\textbf{a})\,(\textbf{b})\,(\textbf{c}) Equal-time ($\tau=0$) two-point correlation~\eqref{eq:zz_corr} for the three models considered, (\textbf{a}) the $\mathbb{Z}_2^3$ model with $\alpha>0$, (\textbf{b}) the $\mathbb{Z}_2^3$ model with $\alpha<0$ and (\textbf{a}) the $\mathbb{Z}_2$ model with $\alpha<0$, with $N=12\times 12$ and inverse temperature $\beta=48$, represented as an intensity map in the Brillouin zone of the triangular lattice. A representative value of $\alpha$ has been chosen for each model. Intensity peaks are visible at specific wave vector $\boldsymbol{q}_{\mathbf{order}}$ reported in Table~\ref{tab:transitions}, and $C(\boldsymbol{q}_{\mathbf{order}},0)$ serves as the definition for a potential order parameter (squared), studied versus $\alpha$. (\textbf{d})\,(\textbf{e})\,(\textbf{f}) Finite-size scaling analysis of $C(\boldsymbol{q}_{\mathbf{order}},0)$ for (\textbf{d}) the $\mathbb{Z}_2^3$ model with $\alpha>0$, (\textbf{e}) the $\mathbb{Z}_2^3$ model with $\alpha<0$ and (\textbf{f}) the $\mathbb{Z}_2$ model with $\alpha<0$. For a second order phase transition it is expected to follow the finite-size critical scaling of Eq.~\eqref{eq:order_scaling}. In the three different cases, we find a single crossing point of the different system sizes with $2\beta/\nu\simeq 1$, as reported in Table~\ref{tab:transitions} (see Appendix~\ref{app:transitions} for information on how we estimated $2\beta/\nu$ and $\alpha_\mathrm{c}$). The data corresponds to the average over the different symmetry-related $\boldsymbol{q}_\mathbf{order}$ wave vectors.}
	\label{fig:order_scaling_sq}
\end{figure*}

\begin{figure}[!t]
	\center
	\includegraphics[width=1\columnwidth,clip]{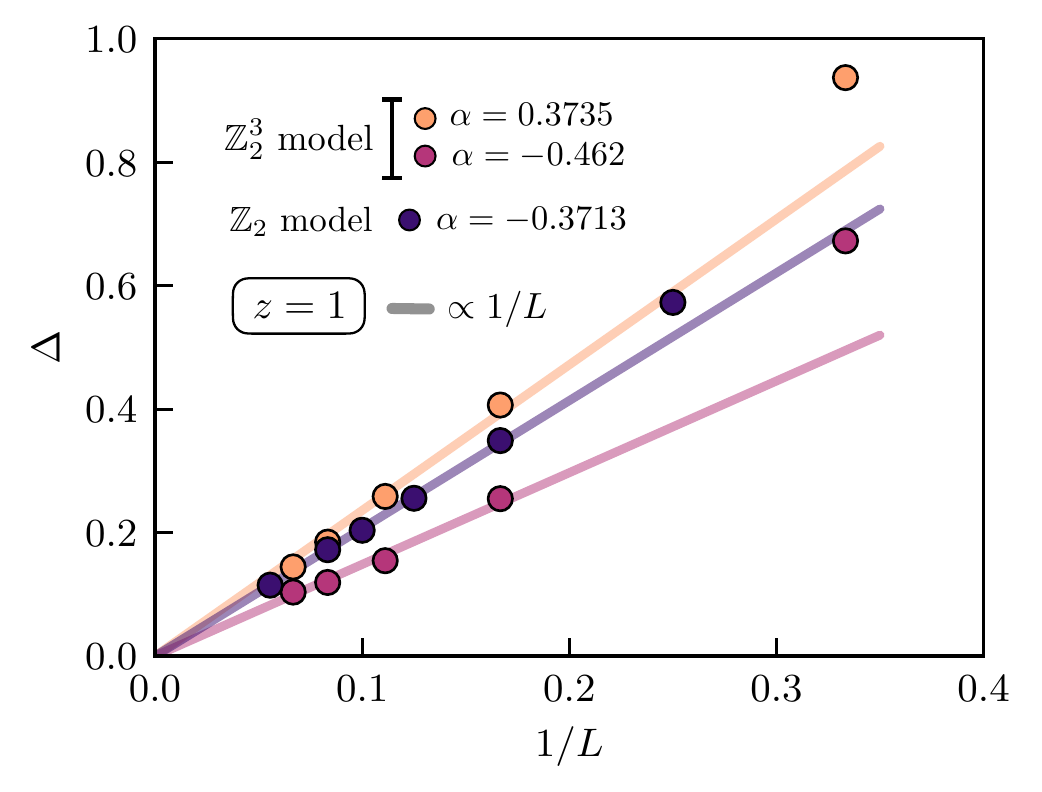}
	\caption{For the three models considered, $\Delta$ gap versus inverse linear system size $L$ at $\alpha\simeq\alpha_\mathrm{c}$, extracted by an exponential fit of the imaginary-time correlation function~\eqref{eq:zz_corr}. For a second order phase transition, it is expected to scale according to Eq.~\eqref{eq:gap_crit}. For the three models, the linear scaling observed versus $1/L$ is compatible with a dynamical exponent $z=1$, when excluding the smallest system size.}
	\label{fig:gap_vs_L}
\end{figure}

There are a total of four cases to be investigated since there are two different symmetry groups ($\mathbb{Z}_2$ and  $\mathbb{Z}^3_2$) and $\alpha\lessgtr 0$. A quantum phase transition is expected in all cases with either a single transition point separating the two phases or a two-step transition via an intermediate phase breaking the protecting symmetry. We first turn our attention to the nature of the transition for the $\mathbb{Z}^3_2$ model at positive and negative $\alpha$ as well as the $\mathbb{Z}_2$ model for $\alpha<0$. The remaining case of the $\mathbb{Z}_2$ model with $\alpha>0$ has already been thoroughly studied by the same authors~\cite{dupont2020}. For completeness, we provide a brief account of its physical properties.

We begin our analysis by probing  the equal-time ($\tau=0$) two-point correlation function~\eqref{eq:zz_corr}, depicted in Figs.~\ref{fig:order_scaling_sq}\,(a,\,b,\,c) for $12\times 12$ lattices, at representative values of $\alpha$, at which order sets in. In all cases, we observe a clear maximum of intensity at specific wave vectors $\boldsymbol{q}_\mathbf{order}$ indicating the presence of long-range magnetic order, as reported in Table~\ref{tab:transitions}. The maximum of intensity serves as the definition for the order parameter (squared) that can be systematically analyzed versus system size and $\alpha$. For a continuous phase transition, the following finite-size critical scaling is expected~\cite{sachdev2001,vojta2003},
\begin{equation}
    C\bigl(\boldsymbol{q}_\mathbf{order},0\bigr)\times L^{2\beta/\nu} = \mathcal{G}\left[L^{1/\nu}\left(\alpha-\alpha_\mathrm{c}\right)\right],
    \label{eq:order_scaling}
\end{equation}
with $\beta$ being the order parameter critical exponent, $\nu$ the correlation length critical exponent, $L$ the linear system size, $\alpha_\mathrm{c}$ the critical point, and $\mathcal{G}$ a universal scaling function. Based on Eq.~\eqref{eq:order_scaling} and as explained in Appendix~\ref{app:transitions}, we determine the position of the critical points $\alpha_\mathrm{c}$ and critical exponents $2\beta/\nu$. The results of this analysis are shown in Fig.~\ref{fig:order_scaling_sq}\,(d,\,e,\,f). Indeed, we find that after rescaling curves corresponding to different systems sizes cross at a single point. For all the three cases considered we estimate $2\beta/\nu\simeq 1$.

We now turn our attention to estimating the gap $\Delta$ between the two $\mathbb{Z}_2$ Ising symmetry sectors by a numerical fit of the imaginary-time correlation function to Eq.~\eqref{eq:zz_corr} at $\boldsymbol{q}_\mathbf{order}$ close to criticality $\alpha\simeq\alpha_\mathrm{c}$. The relevant data set is shown in Appendix~\ref{app:imag_time}. For a continuous transition, the expected finite-size scaling of the gap follows the form~\cite{sachdev2001,vojta2003},
\begin{equation}
    \Delta\bigl(L,\alpha_\mathrm{c}\bigr)\sim L^{-z},
    \label{eq:gap_crit}
\end{equation}
with $z$ being the dynamical critical exponent. In Fig.~\ref{fig:gap_vs_L}, we show the gap versus $1/L$, which displays a linear scaling, compatible with $z=1$ for each of the models.

In principle, we can independently extract the correlation length exponent $\nu$ by rescaling the $x$ axis of Fig.~\ref{fig:order_scaling_sq}\,(d,\,e,\,f) using the scaling argument $\alpha\to L^{1/\nu}(\alpha-\alpha_\mathrm{c})$. Doing so is expected to result in a curve collapse associated with the scaling function $\mathcal{G}$ in Eq.~\eqref{eq:order_scaling}. However, the limited system sizes accessible numerically do not allow for a reliable estimation of $\nu$. Thus, we cannot confidently determine the precise universality classes of each phase transition. In the following, we discuss several possible scenarios describing the various phase transitions, based on our numerical observations and symmetry arguments.

\begin{table}[t]
	\begin{minipage}{0.95\columnwidth}
		\center
		\begin{ruledtabular}
			\begin{tabular}{lccc}
				\thead{\textbf{Model}} & \thead{$\mathbb{Z}_2^3~(\alpha>0)$} & \thead{$\mathbb{Z}_2^3~(\alpha<0)$} & \thead{$\mathbb{Z}_2$ $(\alpha<0)$}\\
				\hline\\[-0.8em]
				\makecell{$\boldsymbol{q}_\mathbf{order}$} & \makecell{\footnotesize $(0,0)$\\ $\left(\frac{4\pi}{3}, 0\right)$\\ $\left(\frac{2\pi}{3}, \frac{2\pi}{\sqrt{3}}\right)$} & \makecell{\footnotesize $\left(\pm \frac{2\pi}{3},\pm \frac{2\pi}{3\sqrt{3}}\right)$\\ $\left(\pm \frac{2\pi}{3},\mp \frac{2\pi}{3\sqrt{3}}\right)$\\ $\left(0,\pm\frac{4\pi}{3\sqrt{3}}\right)$} & \makecell{\footnotesize $\left(0, \frac{2\pi}{\sqrt{3}}\right)$\\ $\left(\pi,\pm \frac{\pi}{\sqrt{3}}\right)$}\\[0.3em]
				\hline\\[-0.8em]
				\makecell{Order type} & \makecell{FM$^3$} & \makecell{$\left(\sqrt{3}\times\sqrt{3}\right)^3$} & \makecell{Commensurate\\stripe}\\[0.3em]
				\hline\\[-0.8em]
				\makecell{$\alpha_\mathrm{c}$} & \makecell{$0.375(5)$} & \makecell{$-0.467(2)$} & \makecell{$-0.374(2)$}\\[0.3em]
				\hline\\[-0.8em]
				\makecell{$2\beta/\nu$} & \makecell{$1.00(8)$} & \makecell{$0.9(1)$} & \makecell{$1.02(2)$}\\[0.3em]
				\hline\\[-0.8em]
				\makecell{$z$} & \makecell{$1$} & \makecell{$1$} & \makecell{$1$}\\[0.3em]
			\end{tabular}
		\end{ruledtabular}
	\end{minipage}
	\caption{The first row indicates the wave vector $\boldsymbol{q}_\mathbf{order}$ at which magnetic order settles in for the different cases, see Fig.~\ref{fig:order_scaling_sq}. The next second row labels the different ordered phases. The third row corresponds to our estimates of the position of the transition point $\alpha_\mathrm{c}$ for each model. The next row reports on the value of the combined critical exponents $2\beta/\nu$ obtained from the finite-size scaling analysis of the numerical data. See Appendix~\ref{app:transitions} for information on how we estimated $2\beta/\nu$ and $\alpha_\mathrm{c}$. The last row shows the dynamical exponent $z$ obtained by the closing of the finite-size gap at criticality.}
	\label{tab:transitions}
\end{table}

\label{sec:order}

\subsection{The $\mathbb{Z}_2^3$ model with $\alpha>0$}

\begin{figure}[!ht]
	\center
	\includegraphics[width=0.6\columnwidth,clip]{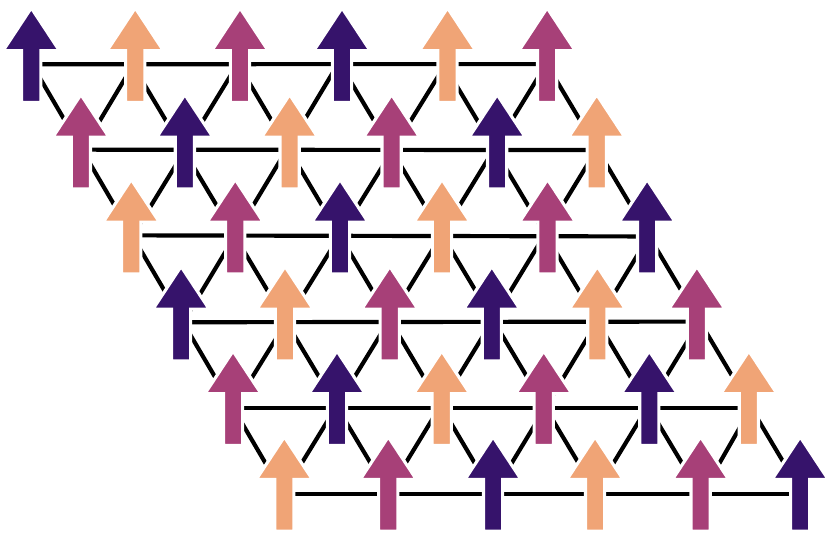}
	\caption{Magnetization for one of the ground states of the FM$^3$ phase, which appears for the $\mathbb{Z}_2^3$ model for $\alpha>0$. The other ground states are obtained by flipping all the spins on one given sublattice (each sublattice is shown in a different color).}
	\label{fig:z23_order_apos}
\end{figure}

The peaks at $\boldsymbol{q}_\mathbf{order}$ in the structure factor, see Table~\ref{tab:transitions}, are consistent with a ferromagnetic phase within each sublattice $\mathsf{A}$, $\mathsf{B}$, and $\mathsf{C}$ as displayed in Fig.~\ref{fig:z23_order_apos} (see also Appendix~\ref{app:bind_ratio}). One way to understand the emergence of this phase is to study the Hamiltonian at $\alpha=1/2$. At that value, it is easy to see that certain spin flips become strictly disallowed, thereby creating a kinetic constraint. The disallowed spin flips are the ones that would change the parity of $N_\mathrm{\triangle_{---}}$ (the number of triangles with three $-1$ spins). The magnetic order close to $\alpha=1/2$ should, therefore, try to maximize the number of flippable spins. It is easy to see that configurations for which each sublattice forms a perfect ferromagnet have all of their spins flippable.

Due to the $\mathbb{Z}_2^3$ symmetry, there is no direct coupling (e.g., $\sigma^z_\mathsf{A}\sigma^z_\mathsf{B}$) between the FM order parameters on each sublattice: They are only coupled through their energy density (like in the classical Ashkin-Teller model~\cite{PhysRev.64.178} in the case of $\mathbb{Z}_2^2$). If this energy density coupling were irrelevant, we would expect three decoupled Ising critical points. However, renormalization group calculations show that this coupling is relevant, and drives the transition either to the $O(N=3)$ universality if $N<N_c$, or to the case of $O(N=3)$ with cubic anisotropy if $N>N_c$ ~\cite{PhysRevB.8.4270,PhysRevB.61.15136,ADZHEMYAN2019332}. 
One would need to access larger system sizes to verify this scenario by extracting the critical exponents $\nu$ and $\beta$ with high precision.

\subsection{The $\mathbb{Z}_2^3$ model with $\alpha<0$}

\footnotetext[1]{One could also imagine a scenario in which this order only appears on one or two sublattices, while the other ones remain paramagnetic, but this scenario seems unlikely.}

\begin{figure}[!t]
	\center
	\includegraphics[width=0.8\columnwidth,clip]{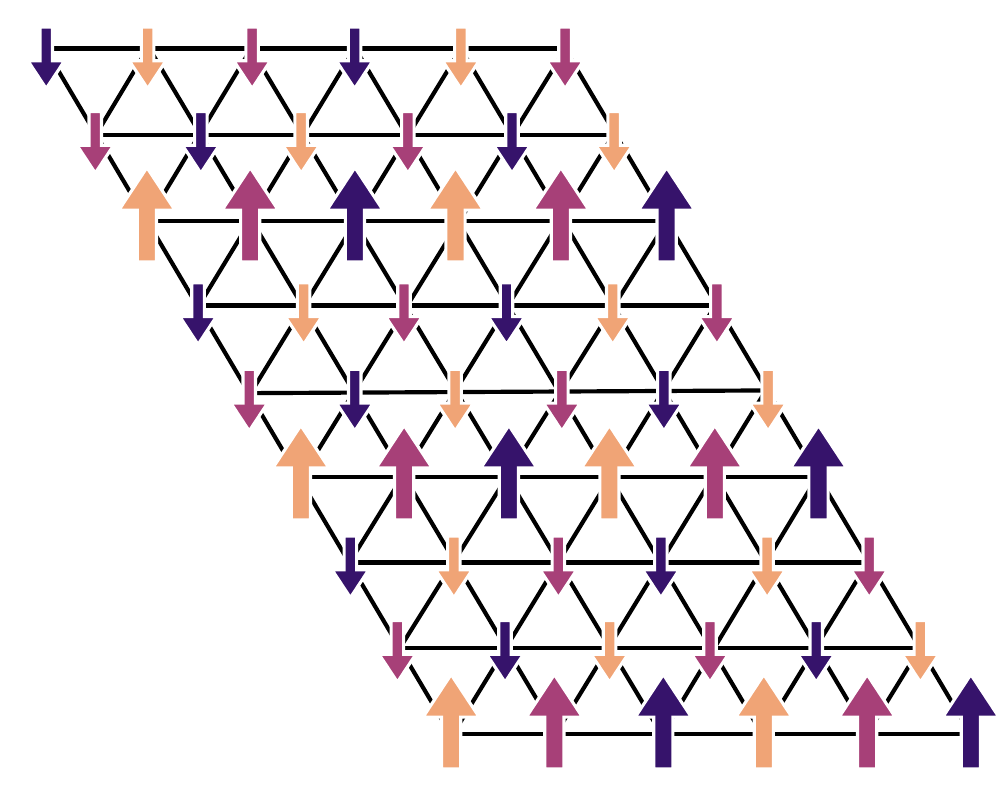}
	\caption{Magnetization for one of the ground states of the $(\sqrt{3}\times\sqrt{3})^3$ phase, which appears for the $\mathbb{Z}_2^3$ model with $\alpha<0$. The other ground states are obtained by flipping all the spins on one given sublattice (each sublattice is shown in a different color). Each sublattice is itself subdivided into three subsublattices (e.g., sublattice $\mathsf{A}$ is divided into $\mathsf{AA}$, $\mathsf{AB}$, and $\mathsf{AC}$), and the magnetization takes a pattern of the type $(1,-\frac{1}{2},-\frac{1}{2})$ for these three subsublattices ($1$ is represented as a big arrow while $-\frac{1}{2}$ is represented as a small one). For the particular ground state shown in the figure, the magnetization has a simple stripe pattern, but this is not the case for all ground states. }
	\label{fig:z23_order_aneg}
\end{figure}

The peaks at $\boldsymbol{q}_\mathbf{order}$ in the structure factor, see Table~\ref{tab:transitions}, are consistent with a $(\sqrt{3}\times\sqrt{3})$ order within each sublattice~\cite{PhysRevLett.38.977,Domany1978,Note1}, see Fig.~\ref{fig:z23_order_aneg}. This phase breaks translation and the $\mathbb{Z}_3$ rotation symmetry of the triangular lattice on top of the Ising spin flip symmetries. Assuming we can neglect the coupling between the sublattices, we would find three independent transitions in the $\mathrm{XY}$ class with $\mathbb{Z}_6$ anisotropy~\cite{Domany1978}. Interestingly, the $\mathbb{Z}_6$ anisotropy is predicted to be dangerously irrelevant, leading to an emergent $\mathrm{U}(1)$ symmetry in the ordered phase below a length scale which diverges as a power of the correlation length~\cite{Balents}. However, the different sublattices are, in reality, coupled and the impact of this coupling should be studied in future work.

\subsection{The $\mathbb{Z}_2$ model with $\alpha<0$}

\begin{figure}[!t]
	\center
	\includegraphics[width=0.6\columnwidth,clip]{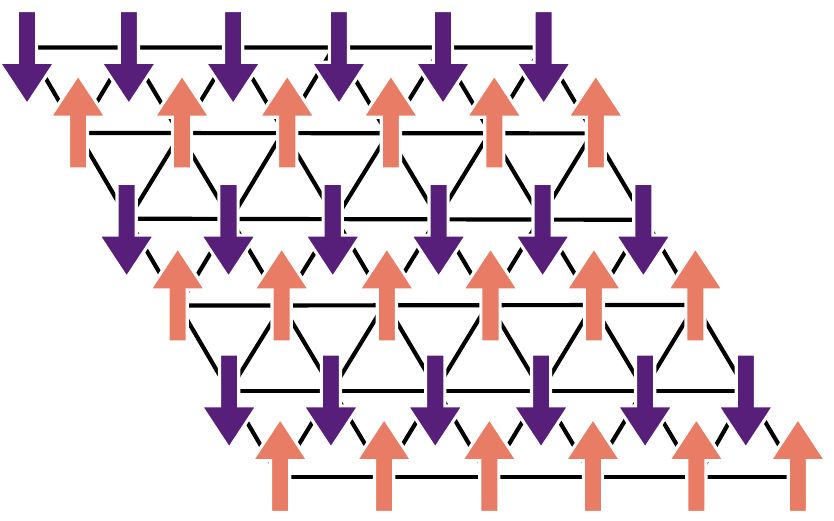}
	\caption{Magnetization for one of the ground states with commensurate stripe order, which appears for the $\mathbb{Z}_2$ model for $\alpha<0$.}
	\label{fig:lg_order_aneg}
\end{figure}

The ordered phase is a stripe phase that breaks translation and the $\mathbb{Z}_3$ rotation symmetry of the triangular lattice on top of the $\mathbb{Z}_2$ Ising spin flip symmetry, see Fig.~\ref{fig:lg_order_aneg}. At $\alpha=-1/2$, certain spin flips become disallowed: the ones that would not change the parity of the number of domain walls. It is easy to see that a stripe phase of a period $2$ has all of its spins in a flippable configuration, which is expected from an energetic point of view.

Based on symmetry alone, a simultaneous breaking of both Ising and the rotational $\mathbb{Z}_3$ symmetry must occur via a first-order transition~\cite{PhysRevB.72.144417,PhysRevLett.116.197201}. In order to reconcile this scenario with our evidence for a second order phase transition, one would need to invoke a weakly first order transition, with a correlation length (finite at the transition) that is larger than the system sizes numerically available. 
Another possibility is to break $\mathbb{Z}_2$ and $\mathbb{Z}_3$ through two consecutive transitions, with an intermediate nematic phase~\cite{PhysRevB.72.144417,PhysRevLett.116.197201}. In that case, the $\mathbb{Z}_2$ breaking transition would be expected to be in the Ising class and would correspond to the transition we observe. Further work would be needed to distinguish these two scenarios.

\subsection{The $\mathbb{Z}_2$ model with $\alpha>0$}

The $\mathbb{Z}_2$ model with $\alpha>0$ was thoroughly investigated in Ref.~\onlinecite{dupont2020}, using the same quantum Monte Carlo methods (SSE and projective) that have been developed in this paper. We provide a brief overview of the main results for completeness.

Similarly to what we have uncovered in this work, there is also an intermediate phase that spontaneously breaks the protecting Ising $\mathbb{Z}_2$ symmetry, and which displays stripe order around the wavevector $|\boldsymbol{q}_{\mathbf{order}}|\simeq 2\pi/5$. A jump in the order parameter at $\alpha_{\mathrm{c}}\approx -0.48$ suggests a first order transition, in agreement with a symmetry-based Ginzburg-Landau analysis prohibiting a continuous transition for the corresponding $\boldsymbol{q}_\mathbf{order}$~\cite{Domany1978}. Remarkably, while one might have expected the intermediate phase to be gapped and confined, it was found to be gapless and dual to a deconfined $\mathrm{U}(1)$ gauge theory due to the incommensurability of the stripe pattern, providing one of the first observations of the ``Cantor deconfinement'' scenario in a microscopic model~\cite{PhysRevLett.64.92,fradkin2004,papanikolaou2007,schlittler2015,PhysRevLett.125.257204}.

\section{Conclusion and perspectives}
\label{sec:conclusion}

Employing numerical simulations based on stochastic series expansion quantum Monte Carlo, we have studied the quantum phase diagram of two Hamiltonians interpolating between trivial and nontrivial paramagnets, protected by $\mathbb{Z}_2$ and $\mathbb{Z}_2^3$ symmetries, respectively. In all cases, we find that the transition happens via an intermediate symmetry-breaking phase, where the protecting symmetry is spontaneously broken, displaying long-range magnetic order. By performing a finite-size scaling analysis of the order parameter, we precisely determined the location of the critical points. The phase diagram of the various models that were investigated in this work are summarized in Fig.~\ref{fig:phase_diagram}. Moreover, we computed the gap $\Delta$ between the two $\mathbb{Z}_2$ Ising symmetry sectors at criticality, and find that it scales as the inverse linear system size of the system, compatible with a dynamical exponent $z=1$. We also discussed the different possible scenarios describing the nature of the phase transitions, which we were not able to single out numerically in the present study.

Despite the fact that we have developed sign-problem-free algorithms for the models considered, there is no known efficient update for sampling the configuration space such as loop updates or cluster-type updates~\cite{sandvik1999,syljuasen2002,sandvik2003,evertz2003}. Therefore, we can only rely on local updates in the configuration space~\cite{sandvik1992}, limiting the system sizes one can simulate. Accessing larger system sizes is paramount in identifying the exact nature of the transitions taking place in these models, calling for the development of a better-suited quantum Monte Carlo algorithm. One could get inspired by the recent progress made for quantum dimer models on the square lattice, also displaying strong geometrical restrictions~\cite{yan2019}.

An important follow-up to this work would be to add terms which frustrate the different magnetic orders, in order to reach multicritical points at which trivial and topological paramagnetic phases could potentially have a direct transition. For example, in the case of stripe phases, it might be possible to reach a quantum Lifshitz point at which the stripe wavevector goes continuously to zero~\cite{kivelson1988,PhysRevB.69.224416,fradkin2004,Ardonne2004493,PhysRevB.65.024504,fradkin2013,PhysRevB.83.125114,moessner2011quantum}. Another possibility is to reach an instance of deconfined quantum critical points, which have been predicted to occur at the transition between different SPT orders in the presence of continuous symmetries~\cite{Senthil1490,2013arXiv1307.8194B,PhysRevX.7.031051,PhysRevX.7.031052,you2016,he2016,you2018,geraedts2017,bi2019,zeng2020}.

In this work, we have focused on bulk properties of the system and thus used periodic boundary conditions. However, it is important to note that the symmetry of the phase diagram around $\alpha=1/2$ for $\alpha>0$ and around $\alpha=-1/2$ for $\alpha<0$ applies to bulk properties, but not to edge properties. In fact, the transitions from a topological paramagnet to a symmetry-breaking phase is expected to have anomalous edge properties compared to the transition from a trivial paramagnet to the same symmetry-breaking phase. Previous studies on such gapless SPT order, also called symmetry-enriched criticality, has been mostly limited to one dimension~\cite{PhysRevB.83.174409,grover2012quantum,PhysRevB.91.235309,parker2017,parker2018,parker2019,verresen2017,verresen2018,2019arXiv190506969V,PhysRevB.103.L100207} (except for Refs.~\cite{grover2012quantum,PhysRevLett.118.087201,parker2017} which include higher dimensional cases). The four models presented here provide an ideal platform to study these phenomena in higher dimensions.

Finally, the models presented here also provide a way of studying transitions between discrete gauge theories. Whereas the trivial paramagnet is dual to the toric code, the nontrivial SPT phases are dual to the double semion (in the case of $\mathbb{Z}_2$) and the non-Abelian $D_4$ quantum double (in the case of $\mathbb{Z}_2^3$). This gauge description was particularly useful to study the transition between toric code and double semion in the case of $\alpha>0$~\cite{dupont2020}. A generalization of this gauge theory description to the $\mathbb{Z}_2$ case for $\alpha<0$ and to the $\mathbb{Z}_2^3$ case is left for future work.

\begin{acknowledgments}
	We are grateful to F. Alet, N. Bultinck, X. Cao, S. Capponi, E. Fradkin, B. Kang, A. Paramekanti, D. Poilblanc, F. Pollmann, P. Pujol, A. W. Sandvik, R. Vasseur, C. Xu and L. Zou for interesting discussions. M.D. was supported by the U.S. Department of Energy, Office of Science, Office of Basic Energy Sciences, Materials Sciences and Engineering Division under Contract No. DE-AC02-05-CH11231 through the Scientific Discovery through Advanced Computing (SciDAC) program (KC23DAC Topological and Correlated Matter via Tensor Networks and Quantum Monte Carlo). S.G. acknowledges support from the Israel Science Foundation, Grant No. 1686/18. T.S. acknowledges the financial support of the Natural Sciences and Engineering Research Council of Canada (NSERC), in particular the Discovery Grant [RGPIN-2020-05842], the Accelerator Supplement [RGPAS-2020-00060], and the Discovery Launch Supplement [DGECR-2020-00222]. This research used the Lawrencium computational cluster resource provided by the IT Division at the Lawrence Berkeley National Laboratory (Supported by the Director, Office of Science, Office of Basic Energy Sciences, of the U.S. Department of Energy under Contract No. DE-AC02-05CH11231). This research also used resources of the National Energy Research Scientific Computing Center (NERSC), a U.S. Department of Energy Office of Science User Facility operated under Contract No. DE-AC02-05CH11231. T.S. contributed to this work prior to joining Amazon.
\end{acknowledgments}

\appendix

\section{Practical information regarding the SSE quantum Monte Carlo algorithm}
\label{app:qmc_sse}

\subsection{Monte Carlo updates}

We first discuss the two types of Monte Carlo moves that we have implemented in order to sample the configuration space. They are called the identity and spin-flip updates.

\subsubsection{Identity update}

For convenience, we also assign a real space position index $j$ to identity operators. The first update is a change of real space position $j$ of such an operator to another real space position $i$: $I_j(p)\rightarrow I_i(p)$. It can be performed as follows,
\begin{enumerate}
	\item Run a loop over each operator of the sequence $S_M$ of the current configuration.
	\item If the operator $O(p)$ is not an identity operator, we move to the next $p$ index.
	\item If the operator at position $p$ is an identity, we get the site $j$ on which it acts on.
	\item We then select at random a site $i\in[1, N]$.
	\item We change with probability one the site on which $I(p)$ is acting from $j$ to $i$.
\end{enumerate}
Basically, this move should always be accepted since it does not change the configuration. Indeed, in the definition~\eqref{eq:partition_function} of the partition function, the identity operators are not specifically associated to a lattice site. We only assign them a lattice site in the algorithm because it makes it much easier to deal with them, especially in regards to the other update.

\subsubsection{Spin-flip update}

The second type of update involves two operators at a time, on different positions $p_1$ and $p_2$ in the sequence but at the same position $j$ in real space,
\begin{equation}
    \Bigl[X_j(p_1), X_j(p_2)\Bigr]\longleftrightarrow \Bigl[I_j(p_1), I_j(p_2)\Bigr],
    \label{eq:spinflip_update1}
\end{equation}
and
\begin{equation}
    \Bigl[I_j(p_1), X_j(p_2)\Bigr]\longleftrightarrow \Bigl[X_j(p_1), I_j(p_2)\Bigr].
    \label{eq:spinflip_update2}
\end{equation}
These updates change the configuration and should be accepted or refused fulfilling detailed balance. In between $p_1$ and $p_2$ at the real space position $j$, there can be as many identity operators $I_j$ as we want but no $X_j$ operators, otherwise these updates would lead to nonvalid configurations. Note the ``periodic boundary condition'' along the $p$ axis, as shown in Fig.~\ref{fig:qmc_conf}. This update can be performed as follows:
\begin{enumerate}
	\item Select a random site $j\in[1, N]$. If the number of operators in the sequence attached to the selected site $j$ is smaller than two, cancel the update. Otherwise, continue.
	\item In the list of operators attached to the site $j$, select one of them at random. We note it $O(p_1)$.
	\item Get the number of operators $N_\mathrm{ops}$ acting on site $j$ between $p_1$ and the first operator $X_j$ encountered (the operator at $p_1$ is excluded from the count and the operator $X$ included). If no operator $X$ is encountered before going back to $p_1$, the count runs up to the previous operator to $O(p_1)$.
	\item Select at random with probability $1/N_\mathrm{ops}$ an operator with $p>p_1$ acting on site $j$. The position of this operator is noted $p_2$.
	\item To fulfill detailed balance in the selection of $p_1$ and $p_2$, the probability to select them in the configuration before and after the update should be the same. This is the case in the selection of $p_1$ but the probability to select $p_2$ depends on the nature of the operator at $p_2$. Consider this: If it is an $X$ operator then the probability to select it is $1/N_\mathrm{ops}$. After an update changing $X$ to $I$, the probability to select the identity operator at $p_2$ will be modified. In the current scheme, the probability to select it would be $1/(N_\mathrm{ops}+N'_\mathrm{ops})$ with $N'_\mathrm{ops}$ the number of operators between $I_j(p_2)$ and the next operator $X$ acting on site $j$. To correct this imbalance in the selection of $p_2$ if $O_j(p_2)\equiv X_j(p_2)$, we cancel this selection with probability, $P_\mathrm{cancel} = 1 - N_\mathrm{ops}/(N_\mathrm{ops}+N'_\mathrm{ops})$.
	\item If the selection is not canceled, we suggest the update according to Eq.~\eqref{eq:spinflip_update2}.
\end{enumerate}

The probability to accept such an update involves the ratio of the weights of the configurations after ``a'' and before ``b'', i.e., $P_\mathrm{accept}=\mathrm{min}(R_{\mathrm{b}\rightarrow\mathrm{a}},1)$, with,
\begin{equation}
    R_{\mathrm{b}\rightarrow\mathrm{a}}=\frac{W\Bigl(|\sigma_\mathrm{a}\rangle, S^\mathrm{a}_M\Bigr)}{W\Bigl(|\sigma_\mathrm{b}\rangle, S^\mathrm{b}_M\Bigr)}.
\end{equation}
Specifically, by defining the ratio of matrix elements
\begin{equation}
    r=\frac{\langle\sigma_\mathrm{a}|\prod_{O\in S^\mathrm{a}_M}O |\sigma_\mathrm{a}\rangle}{\langle\sigma_\mathrm{b}| \prod_{O\in S^\mathrm{b}_M}O|\sigma_\mathrm{b}\rangle}\geq 0,
    \label{eq:ratio_melts}
\end{equation}
one finds that the acceptance probability of the updates~\eqref{eq:spinflip_update1} and~\eqref{eq:spinflip_update2} is given by
\begin{equation}
    \begin{split}
        R_{XX\rightarrow II}&=r\Bigl[(M-n+2)(M-n+1)\Bigr]/\bigl(N\beta\bigr)^2,\\
        R_{II\rightarrow XX}&=r\bigl(N\beta\bigr)^2/\Bigl[(M-n-1)(M-n)\Bigr],\\
        R_{IX\rightarrow XI}&=R_{XI\rightarrow IX}=r,
    \end{split}
\end{equation}
with $n$ the number of nonidentity operators before the update. The $N^2$ factor comes from the fact that we label the identity operators with a lattice site. In practice, the ratio of matrix elements~\eqref{eq:ratio_melts} can be efficiently computed since the update only involves a change of two operators on the same site $j$ at position $p_1$ and $p_2$.

\subsection{Initialization and thermalization}

We initially start with a configuration only involving $M\approx 10$ identity operators, randomly positioned in real space. An initial spin configuration $|\sigma\rangle$ is also generated at random. The thermalization process consists of running consecutively the identity and spin-flip updates and increasing the size of the sequence of operators $M$ by about $10\%$ (by randomly adding identity operators at the end of the sequence) when $n>0.8M$, with $n$ the number of nonidentity operators. This ensures that $n<M$ in the following, when updates are performed in order to get measurements.

\section{Projective quantum Monte Carlo}
\label{app:qmc_proj}

The basic idea of projective quantum Monte Carlo~\cite{becca2017} lies behind the power method,
\begin{equation}
    |\psi_\mathrm{gs}\rangle\sim \lim_{m\to+\infty}\mathcal{H}^m|\phi\rangle,
    \label{eq:power_method}
\end{equation}
with $\langle\psi_\mathrm{gs}|\phi\rangle\neq 0$. This algorithm was used in Ref.~\onlinecite{dupont2020} to study the $\mathbb{Z}_2$ model at $\alpha=1/2$.

\subsection{Configuration space}

Based on Eq.~\eqref{eq:power_method}, we define the following equivalent of the ``partition function'' (or normalization) at order $m$,
\begin{equation}
    \mathcal{Z}(m) = \bigl\langle\phi\bigl|\mathcal{H}^m\mathcal{H}^m\bigl|\phi'\bigr\rangle.
\end{equation}
Choosing for initial state $|\phi\rangle=|\phi'\rangle=\sum_{\{|\sigma\rangle\}}|\sigma\rangle$ and using the same notation as Eq.~\eqref{eq:hamiltonian_qmc} for the Hamiltonian, one arrives to
\begin{equation}
    \mathcal{Z}(m) = \sum_{\{|\sigma\rangle\}}\sum_{\{|\sigma'\rangle\}}\Bigl\langle\sigma\Bigl|\left(\sum\nolimits_j X_j\right)^{2m}\Bigl|\sigma'\Bigr\rangle.
\end{equation}
Expanding the power as the product of all the possible sequences $\{S_{2m}\}$ of operators $X_j$ of length $2m$, one gets
\begin{equation}
    \mathcal{Z}(m) = \sum_{\{|\sigma\rangle\}}\sum_{\{|\sigma'\rangle\}}\sum_{\{S_{2m}\}}\Bigl\langle\sigma\Bigl|\prod_{X_j\in S_{2m}}X_j\Bigl|\sigma'\Bigr\rangle.
    \label{eq:final_zm}
\end{equation}
One can rewrite the partition function~\eqref{eq:final_zm} as
\begin{equation}
    \mathcal{Z}(m)=\sum_{\{|\sigma\rangle\}}\sum_{\{|\sigma'\rangle\}}\sum_{\{S_{2m}\}}W\Bigl(|\sigma\rangle, |\sigma'\rangle, S_{2m}\Bigr),
\end{equation}
where $W(|\sigma\rangle, |\sigma'\rangle, S_{2m})$ is the weight of a configuration with a probability $P(|\sigma\rangle, |\sigma'\rangle, S_{2m})=W(|\sigma\rangle, |\sigma'\rangle, S_{2m})/\mathcal{Z}(m)\geq 0$ for the parameter range $\alpha\in[-0.5,+0.5]$.

\begin{figure}[t]
	\center
	\includegraphics[width=0.6\columnwidth,clip]{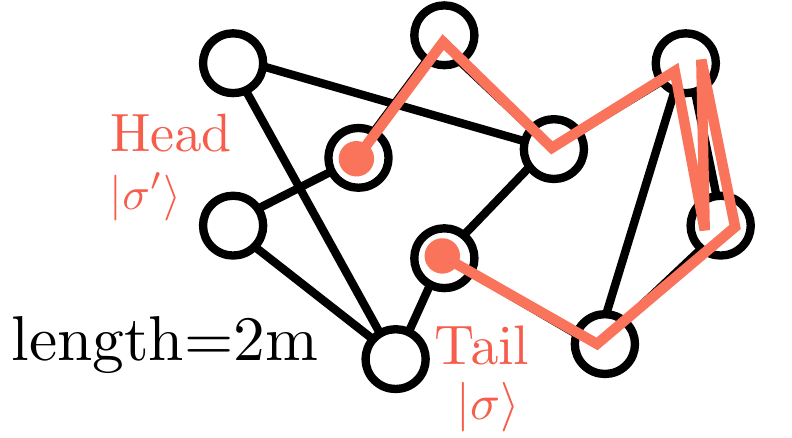}
	\caption{Example of a Monte Carlo configuration in the projective algorithm: It is a snake of length $2m$ in the graph where the vertices are basis states and the edges are single spin flips. The ``head'' and the ``tail'' of the snake are $|\sigma'\rangle$ and $|\sigma\rangle$, respectively. Its ``body'' consists of all intermediate basis states connecting $|\sigma\rangle$ to $|\sigma'\rangle$ by applying the spin flips of the sequence $S_{2m}$.}
	\label{fig:qmc_conf_snake}
\end{figure}

The configurations have a convenient graphical representation: It represents a ``snake'' of length $2m$ on a graph where the vertices are basis states and the edges are single spin flips. The ``head'' and the ``tail'' of the snake are $|\sigma'\rangle$ and $|\sigma\rangle$, respectively. Its ``body'' consists of all intermediate basis states connecting $|\sigma\rangle$ to $|\sigma'\rangle$ by applying the spin flips of the sequence $S_{2m}$. Including $|\sigma\rangle$ and $|\sigma'\rangle$, there are $(2m+1)$ vertices in total, see Fig.~\ref{fig:qmc_conf_snake}.

\begin{figure*}[t]
	\center
	\includegraphics[width=2.0\columnwidth,clip]{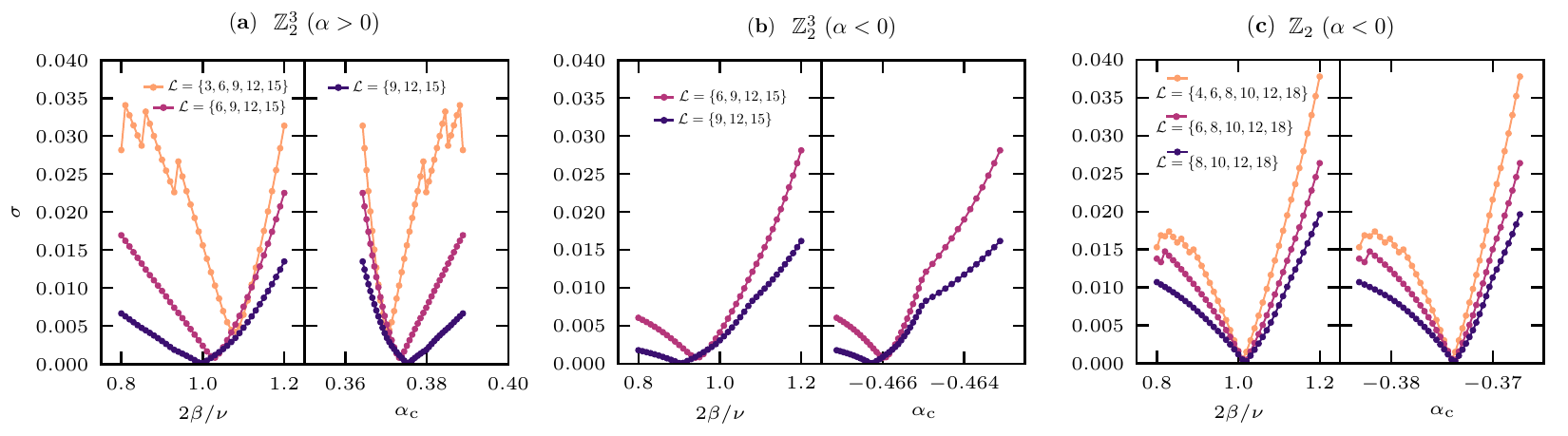}
	\caption{(\textbf{a},\,\textbf{b},\,\textbf{c}) Minimum spreading of the crossing points defined in Eq.~\eqref{eq:crossing} between all pairs of linear system sizes in $\mathcal{L}$, according to Eq.~\eqref{eq:crossings_spreading}. A minimum is observed in all cases with a drift of its position observed as one removes the smallest system sizes from the set $\mathcal{L}$. The values of $2\beta/\nu$ and $\alpha_\mathrm{c}$ reported in Table~\ref{tab:transitions} correspond to the minimum considering the largest system sizes only (violet curves). The error bars that we give reflect the difference with respect to the data set containing all sizes (see text). For the $\mathbb{Z}_2^3$ model at $\alpha<0$, we did not take into account $N=3\times 3$ as its crossing point with other system sizes was always leading to outliers.}
	\label{fig:best_crossings}
\end{figure*}

\subsection{Monte Carlo update}

The update aims at moving the snake of Fig.~\ref{fig:qmc_conf_snake} on the graph: It corresponds to translating the whole snake by one vertex at a time, either by its head or its tail. It can be implemented as follows,
\begin{enumerate}
	\item Select the head or the tail of the snake at random with probability $1/2$.
	\item Independently of which one has been selected, there are $N$ spins in $|\sigma'\rangle$ and $|\sigma\rangle$ which can be potentially flipped. One is selected at random with probability $1/N$. The new basis state obtained when applying the spin flip corresponds to the update proposal, where the head or the tail will move if it is accepted.
\end{enumerate}
The probability to accept such an update involves the ratio of the weights of the configurations after ``a'' and before ``b'' the move, i.e., $P_\mathrm{accept}=\mathrm{min}(R_{\mathrm{b}\rightarrow\mathrm{a}},1)$, with,
\begin{equation}
    R_{\mathrm{b}\rightarrow\mathrm{a}}=\frac{\langle\sigma_\mathrm{a}|\prod_{X_j\in S^\mathrm{a}_{2m}}X_j |\sigma_\mathrm{a}'\rangle}{\langle\sigma_\mathrm{b}| \prod_{X_j\in S^\mathrm{b}_{2m}}X_j|\sigma_\mathrm{b}'\rangle}\geq 0,
\end{equation}
which can be readily computed since only two operators $X_j$ differ between the two sequences $S^\mathrm{b}_{2m}$ and $S^\mathrm{a}_{2m}$. Because a single update is highly local in the configuration space, we perform $2m$ of them consecutively in what we call an actual update for this algorithm.

At $\alpha=\pm 1/2$, some spin flips become impossible (the matrix element is strictly zero), while the ones which remain possible all have the same matrix element $1$. In practice, one can take advantage of this and slightly adapt the above algorithm by only suggesting moves of the head/tail to configurations where the spin is flippable (note that the probability $P_\mathrm{accept}$ needs to be modified accordingly to satisfy detailed balance).

\subsection{Initialization and thermalization}

The initialization and thermalization parts of the algorithms increase the length $\ell$ of the snake until it reaches the desired value $2m$. We typically start with a snake of length $\ell=2$, generated at random on the graph and perform a number of updates of the order of the number of lattice sites (as described above). When this is done, we symmetrically (with respect to the tail and the head) increase the length of the snake $\ell\to\ell+2$. The positions of the new head and tail are selected at random. We then repeat this whole process until $\ell\equiv 2m$. Although the the position of the initial snake and the position of the new head and tail are random, we have to ensure that the configuration is valid by making sure that the corresponding operators $X_j$ introduced in the sequence do not lead to zero matrix elements.

Increasing the length of the snake on the fly allows one to check on whether or not its current size is sufficiently long to probe the ground state or not (by regularly performing measurements, of the energy for instance), and adjust $m$ accordingly.

\subsection{Measurements}

With the projective algorithm, the measurement of an observable $O$ takes the form,
\begin{equation}
    \bigl\langle O\bigr\rangle = \bigl\langle\phi\bigl|\mathcal{H}^mO\mathcal{H}^m\bigl|\phi'\bigr\rangle\bigl/\mathcal{Z}(m).
\end{equation}
From the snake configuration perspective, if $O$ is a diagonal observable in the computational basis, it is measured on the spin configuration positioned in the middle of the snake. If one wants to measure the energy $\langle\mathcal{H}\rangle$, it can be achieved by averaging $-\sum_j\bigl(1-|\alpha|+\alpha \vartheta_j\bigr)$ over the head or tail spin configurations $|\sigma'\rangle$ and $|\sigma\rangle$.

\begin{figure*}[t]
	\center
	\includegraphics[width=2.0\columnwidth,clip]{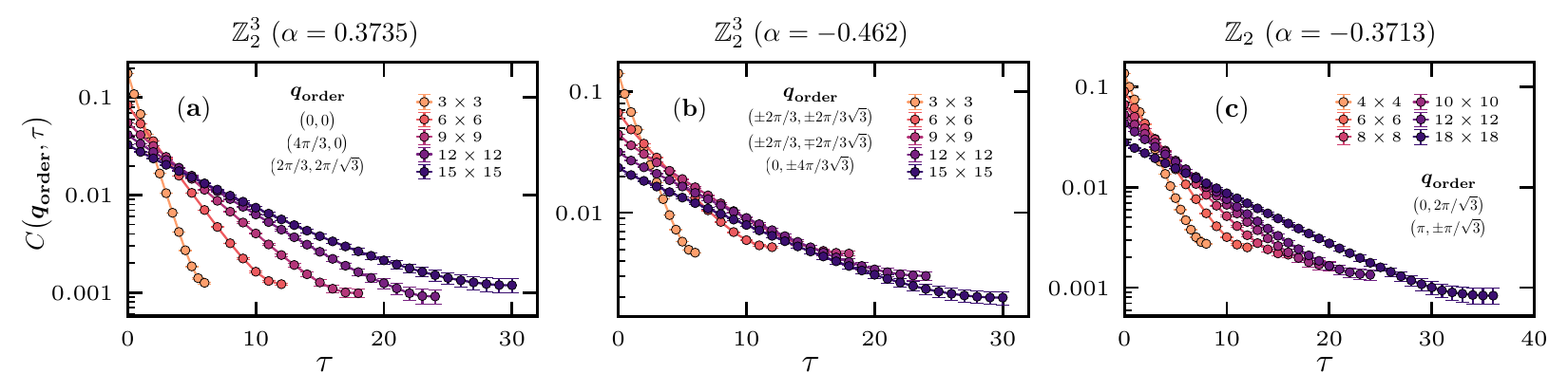}
	\caption{(\textbf{a},\,\textbf{b},\,\textbf{c}) Fourier transform of the imaginary-time two-point correlation function~\eqref{eq:zz_corr} for the $\mathbb{Z}_2^3$ and $\mathbb{Z}_2$ models at $\alpha\simeq\alpha_\mathrm{c}$, computed at $\boldsymbol{q}_\mathbf{order}$, see Table~\ref{tab:transitions}. An exponential fit of the data versus imaginary time gives access to the gap $\Delta$ between the two $\mathbb{Z}_2$ Ising symmetry sectors for a given finite-size system. The value of the gap is reported in Fig.~\ref{fig:gap_vs_L} with a finite-size critical scaling of the form~\eqref{eq:gap_crit}.}
	\label{fig:imag_time}
\end{figure*}

\section{Locating the transitions}
\label{app:transitions}

We assume a second order phase transition for which the following finite-size critical scaling is expected at criticality $\alpha=\alpha_\mathrm{c}$~\cite{sachdev2001,vojta2003} [see also Eq.~\eqref{eq:order_scaling}],
\begin{equation}
    C\bigl(\boldsymbol{q}_\mathbf{order},0\bigr)\times L^{2\beta/\nu}\sim\mathrm{constant},
    \label{eq:crossing}
\end{equation}
with $\beta$ the order parameter critical exponent, $\nu$ the correlation length critical exponent, and $L$ the linear system size. $C\bigl(\boldsymbol{q}_\mathbf{order},0\bigr)$ is measured in quantum Monte Carlo for different values of $\alpha$ and linear system sizes $L$. Both $2\beta/\nu$ and $\alpha_\mathrm{c}$ are unknown. In order to determine them, we set $2\beta/\nu$ as a parameter and find the value for which the crossing of the different system sizes is as close as possible to a single point, which gives $\alpha_\mathrm{c}$.

In practice, we have a set of data corresponding to different system sizes $\mathcal{L}=\{L_1,\,L_2,\,\ldots,\,L_{N_\mathrm{sizes}}\}$. We compute the $\binom{\mathcal{L}}{2}$ possible combinations between them. For a given value of the exponent, and for each pair, we get the coordinates of the crossing point between the two curves $(x_\mathrm{c},\,y_\mathrm{c})$ (we do a linear interpolation between the different data points). From the resulting list of coordinates $\{(x_\mathrm{c},\,y_\mathrm{c})\}$, we quantify their spreading by computing the standard deviation of the euclidean distances $d_\mathrm{c}=\sqrt{x_\mathrm{c}^2 + y_\mathrm{c}^2}$,
\begin{equation}
    \sigma\bigl(2\beta/\nu\bigr) = \sqrt{\overline{d_\mathrm{c}^2} - \overline{d_\mathrm{c}}}\quad\mathrm{for~a~given}~\mathcal{L}.
    \label{eq:crossings_spreading}
\end{equation}
We estimate the best exponent $2\beta/\nu$ from the minimum of $\sigma$ and estimate $\alpha_\mathrm{c}$ as the average over the $x_\mathrm{c}$ coordinates for the corresponding best exponent. This method puts all the system sizes on the same level (the smallest and the largest), but we know that the crossings can exhibit some drifts with the system sizes. To that end, we repeat the procedure by removing from the set $\mathcal{L}$ the smallest system size and the two smallest ones (we are limited on how far we can go by the total size of $\mathcal{L}$).

Results are plotted in Fig.~\ref{fig:best_crossings} for the three models considered, with $\sigma$ showing a well-defined minimum in all cases. The values of $2\beta/\nu$ and $\alpha_\mathrm{c}$ reported in Table~\ref{tab:transitions} correspond to the minimum considering the largest system sizes only (violet curves). The error bars that we give reflect the difference with respect to the data set containing all sizes. In that sense, this is more of an upper bound since we see that the difference between the position of the minima decreases when removing the smallest system sizes.

\begin{figure*}[!t]
	\center
	\includegraphics[width=1.7\columnwidth,clip]{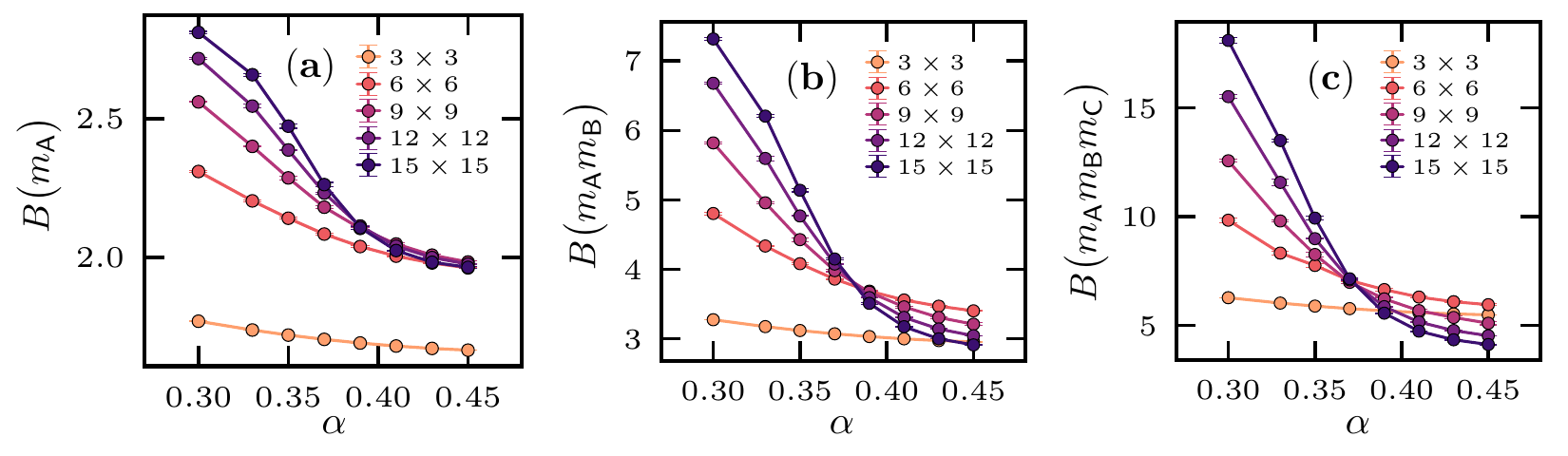}
	\caption{(\textbf{a},\,\textbf{b},\,\textbf{c}) Binder ratio $B$ of Eq.~\eqref{eq:binder_ratio} for the $\mathbb{Z}_2^3$ model with $\alpha<0$ involving the magnetization $m_\mathsf{X}$ of each sublattice $\mathsf{X}\in[\mathsf{A},\mathsf{B},\mathsf{C}]$ as defined in Eq.~\eqref{eq:mag_sublat}.}
	\label{fig:z23_apos_binder}
\end{figure*}

\section{Extracting the gap $\Delta$ between the two Ising symmetry sectors}
\label{app:imag_time}

The gap $\Delta$ between the two $\mathbb{Z}_2$ Ising symmetry sectors, reported in Fig.~\ref{fig:gap_vs_L} for different system sizes $L\times L$, is indirectly accessed in quantum Monte Carlo by performing an exponential fit $\propto\exp(-\Delta\tau)$ of the imaginary-time correlation data displayed in Fig.~\ref{fig:imag_time}. The imaginary time $\tau$ is defined within the range $[0,\beta/2]$ with inverse temperature $\beta=4L$ considered. The fit to extract the gap is only performed over the range showing a genuine exponential decay.

\section{Binder ratio with individual sublattices for the $\mathbb{Z}_2^3$ model with $\alpha>0$}
\label{app:bind_ratio}

For the $\mathbb{Z}_2^3$ model with $\alpha>0$, in order to check whether or not ferromagnetic order settles in one, two, or the three sublattices $\mathsf{A}$, $\mathsf{B}$, and $\mathsf{C}$ (whereas the others remain paramagnetic), see Fig.~\ref{fig:model}\,(c), we compute the following Binder ratio,
\begin{equation}
    B(m)=\frac{\bigl\langle m^4\bigr\rangle}{\bigl\langle m^2\bigr\rangle^2}\quad\mathrm{with}\quad m\equiv
    \begin{cases}
        m_\mathsf{A},\\
        m_\mathsf{A}m_\mathsf{B},\\
        m_\mathsf{A}m_\mathsf{B}m_\mathsf{C},
    \end{cases}
\label{eq:binder_ratio}
\end{equation}
which involves the magnetization of each sublattice independently,
\begin{equation}
    m_\mathsf{A}=\frac{3}{N}\sum\nolimits_{i\in\mathsf{A}}\sigma^z_i.
    \label{eq:mag_sublat}
\end{equation}
It is plotted in Fig.~\ref{fig:z23_apos_binder}\,(a,\,b,\,c) versus $\alpha$, with a crossing of the largest system sizes observed in all cases, meaning that the three sublattices experience long-range ordering. For $B(m_\mathsf{A})$ and $B(m_\mathsf{A}m_\mathsf{B})$, the crossing point seems to happen at a larger $\alpha$ value than for the structure factor $C(\boldsymbol{q}_{\mathbf{order}},0)$ of the main text. However, this is attributed to the ``effectively'' smaller system size when considering the sublattices independently (each account for $N/3$ lattice sites only), as compared to the other Binder ratio $B(m_\mathsf{A}m_\mathsf{B}m_\mathsf{C})$ or the structure factor since we expect that $m_\mathsf{A}^2\equiv C(\boldsymbol{0},0)$.

\bibliography{bibliography}

\end{document}